\documentclass[superscriptaddress,twocolumn,showpacs,pra,floatfix]{revtex4}

\usepackage{epsfig}
\usepackage{color}
\usepackage{tabularx}
\usepackage{epsfig}
\usepackage{amsmath}
\usepackage{amssymb}
\usepackage{graphicx}
\usepackage{wasysym}
\usepackage{dcolumn}
\usepackage{bm}
\usepackage{psfrag}
\usepackage{times}

\begin{document}

\title{Machine learning in physics: The pitfalls of poisoned training sets}

\author{Chao Fang}
\affiliation{Department of Physics and Astronomy, Texas A$\&M$ University,
College Station, Texas 77843-4242, USA}

\author{Amin Barzeger}
\affiliation{Department of Physics and Astronomy, Texas A$\&M$ University, 
College Station, Texas 77843-4242, USA}
\affiliation{Microsoft Quantum, Microsoft, Redmond, WA 98052, USA}

\author{Helmut G.~Katzgraber}
\affiliation{Microsoft Quantum, Microsoft, Redmond, WA 98052, USA}

\begin{abstract} 

Known for their ability to identify hidden patterns in data, artificial
neural networks are among the most powerful machine learning tools.
Most notably, neural networks have played a central role in identifying
states of matter and phase transitions across condensed matter physics.
To date, most studies have focused on systems where different phases of
matter and their phase transitions are known, and thus the performance of
neural networks is well controlled.  While neural networks present an
exciting new tool to detect new phases of matter, here we demonstrate
that when the training sets are poisoned (i.e., poor training data or
mislabeled data) it is easy for neural networks to make misleading
predictions.

\end{abstract}


\maketitle

\section{Introduction}

Machine learning methods \cite{haykin:08,goodfellow:16,bishop:06} have
found applications in condensed matter physics detecting phases of
matter and transitions between these on both quantum and classical
systems (see, for example,
Refs.~\cite{ronhovde:11,nussinov:16,carrasquilla:17,chng:17,tanaka:17,kashiwa:18x}).
Different approaches exist, such as lasso
\cite{santosa:86,tibshirani:94}, sparse regression
\cite{mateos:10,candela:05}, classification and regression trees
\cite{rokach:14,shalev:14,mehta:02}, as well as boosting and support
vector machines \cite{james:13,hsu:10,platt:99,widodo:07,joachims:98}.
Neural networks \cite{lecun:98,zhang:90} are the most versatile and
powerful tools, which is why they are commonly used in scientific
applications.

Convolutional neural networks (CNNs), in particular, are specialized
neural networks for processing data with a grid-like topology. Familiar
examples include time-series data, where samples are taken in intervals,
and images (two-dimensional data sets).  The primary difference between
neural networks and convolutional neural networks lies in how hidden
layers are managed. In CNNs, a {\em convolution} is applied to divide
the feature space into smaller sections emphasizing local trends.
Because of this, CNNs are ideally-suited to study physical models on
hypercubic lattices.  Recently, it was demonstrated that CNNs can be
applied to the detection of phase transitions in Edwards-Anderson Ising
spin glasses on cubic lattices \cite{munoz:19x}. It was shown that the
critical behavior of a spin glass with bimodal disorder can be inferred
by training the model using data that has Gaussian interactions between
the spins. The use of CNNs also results in a reduced numerical effort,
which means one could potentially access larger system sizes often
needed to overcome corrections to scaling in numerical studies.  As
such, pairing specialized hardware to simulate Ising systems
\cite{alvarez:10a,banos:12,baity:14} with machine learning techniques
might one day elucidate properties of spin glasses and related systems.
However, as we show in this work, the use of poor input data can result
in erroneous or even unphysical results. This (here inadvertent) {\em
poisoning} of the training set is well known in computer science where
small amounts of bad data can strongly affect the accuracy of neural
network systems. For example, Steinhardt {\em et
al.}~\cite{steinhardt:17x} demonstrated that already small amounts of
bad data can result in a sizable drop in the classification accuracy.
References \cite{jagielsky:18,alfeld:16,shi:19} furthermore demonstrate
that data poisoning can have a strong effect in machine learning.
Reference \cite{jiang:19} focuses on adversarial manipulations
\cite{nelson:08,newell:14} of simulational and experimental data in
condensed matter physics applications. In particular, they show that
changing individual variables (e.g., a pixel in a data set) can generate
misleading predictions  This suggests that results from machine learning
algorithms sensitively rely on the quality of the training input.

In this work, we demonstrate that the use of poorly-thermalized Monte
Carlo data or simply mislabeled data can result in erroneous estimates
of the critical temperatures of Ising spin-glass systems. As such, we
focus less on adversarial cases, but more on accidental cases of poor
data preparation.  We train a CNN with data from a Gaussian Ising spin
glass in three space dimensions and then use data generated for a
bimodal Ising spin glass to predict the transition temperature of the
same model system, albeit with different disorder. In addition, going
beyond the work presented in Ref.~\cite{jiang:19}, we introduce an
analysis pipeline that allows for the precise determination of the
critical temperature.  While good data results in a relatively accurate
prediction, the use of poorly-thermalized or mislabeled data produce
misleading results. This should serve as a cautionary tale when using
machine learning techniques for physics applications.

The paper is structured as follows. In Sec.~\ref{sec:mod} we introduce
the model used in the study, as well as simulation parameters for both
training and prediction data. In addition, we outline the implementation
of the CNN as well as the approach used to extract the
thermodynamic critical temperature, followed by results and concluding
remarks.

\section{Model and numerical details}
\label{sec:mod}

To illustrate the effects of poisoned training sets we study the
three-dimensional Edwards-Anderson Ising spin glass
\cite{edwards:75,binder:86,mezard:87,young:98,stein:13} with a neural
network implemented in TensorFlow \cite{albadi:16}. The model
is described by the Hamiltonian
\begin{equation}
{\mathcal H}=-\sum_{\langle i,j \rangle} J_{ij} s_i s_j ,
\end{equation}
where each $J_{ij}$ is a random variable drawn from a given
symmetric probability distribution, either bimodal, i.e.,
$\pm 1$ with equal probability, or Gaussian with zero 
mean and unit variance. In addition, $s_i = \pm 1$
represent Ising spins, and the sum is over nearest neighbors on a cubic
lattice with $N$ sites.

Because spin glasses do not exhibit spatial order below the spin-glass
transition, we measure the site-dependent spin overlap \cite{ sherrington:75, parisi:80, parisi:83}
\begin{equation}
\label{overlaps}
q_i = S^{\alpha}_{i}S^{\beta}_{i},
\end{equation}
between replicas $\alpha$ and $\beta$. In the overlap space, the system
is reminiscent of an Ising ferromagnet, i.e., approaches for ferromagnetic
systems introduced in
Refs.~\cite{carrasquilla:17,chng:17} can be used. For low temperatures, $q
= (1/N)\sum_i q_i \to 1$, whereas for $T \to \infty$, $q \to 0$.  For an
infinite system, $q$ abruptly drops to zero at the critical temperature
$T_c$. Therefore, the overlap space is well suited to detect the
existence of a phase transition in a disordered system, even beyond
spin glasses. In the overlap space, the spin-glass phase transition can
be visually seen as the formation of disjoint islands with identical spin configurations. 
As such, the problem of phase identification in physical systems is 
reminiscent of an image classification problem where CNN's are
shown to be highly efficient compared to fully-connected neural 
networks (FCN).

\subsection{Data generation}

We use parallel tempering Monte Carlo \cite{hukushima:96} to generate
configurational overlaps. Details about the parameters used in the Monte
Carlo simulations are listed in Tab.~\ref{tab:train} for the training
data with Gaussian disorder. The parameters for the prediction data with
bimodal disorder are listed in Tab.~\ref{tab:test}.

\begin{table}[h]
\caption{
Parameters for the training samples with Gaussian disorder. $L$ is the
linear size of a system with $N = L^3$ spins, $N_{\rm sa}$ is the number of
samples, $N_{\rm sw}$ is the number of Monte Carlo sweeps for each of
the replicas for a single sample, $T_{\rm min}$ and $T_{\rm max}$ are the lowest and
highest temperatures simulated, $N_{T}$ is the number of temperatures
used in the parallel tempering Monte Carlo method for each system size
$L$, and $N_{\rm con}$ is the number of configurational overlaps for a given
temperature in each instance.  \label{tab:train} }
\begin{tabular*}{\columnwidth}{@{\extracolsep{\fill}}c r r r r r l }
\hline
\hline
$L$ & $N_{\rm sa}$ & $N_{\rm sw}$ & $T_{\rm min}$ & $T_{\rm max}$ & $N_{T}$ & $N_{\rm con}$ \\
\hline
$8$ & $20000$ & $50000$ & $0.80$ &$1.21$ &$20$ & $100$\\
\hline
$10$ & $10000$ & $40000$ &  $0.80$ &$1.21$  &$20$ & $100$\\
\hline
$12$ & $20000$ & $655360$ &  $0.80$ &$1.21$ &$20$ & $100$\\
\hline
$14$ & $10000$ & $1050000$ &  $0.80$ &$1.21$ &$20$ & $100$\\
\hline
$16$ & $5000$ & $1050000$ &  $0.80$ &$1.21$ &$20$ & $100$\\
\hline
\hline
\end{tabular*}
\end{table}

\begin{table}[h]
\caption{
Parameter for the prediction samples with bimodal disorder. $L$ is the
linear size of the system, $N_{\rm sa}$ is the number of samples, $N_{\rm sw}$ is
the number of Monte Carlo sweeps for each of the replicas of a single
sample, $T_{\rm min}$ and $T_{\rm max}$ are the lowest and highest temperatures
simulated, $N_{T}$ is the temperature numbers used in parallel tempering
method for each linear system size $L$, and $N_{\rm con}$ is the number of
configurational overlaps for a given temperature in each instance.
\label{tab:test} }
\begin{tabular*}{\columnwidth}{@{\extracolsep{\fill}}c r r r r r l }
\hline
\hline
$L$ & $N_{\rm sa}$ & $N_{\rm sw}$ & $T_{\rm min}$ & $T_{\rm max}$ & $N_{T}$ & $N_{\rm con}$ \\
\hline
$8$ & $15000$ & $80000$ & $1.05$ &$1.25$ &$12$ & $500$\\
\hline
$10$ & $10000$ & $300000$ &  $1.05$ &$1.25$  &$12$ & $500$\\
\hline
$12$ & $4000$ & $300000$ &  $1.05$ &$1.25$ &$12$ & $500$\\
\hline
$14$ & $4000$ & $1280000$ &  $1.05$ &$1.25$ &$12$ & $500$\\
\hline
$16$ & $4000$ & $1280000$ &  $1.05$ &$1.25$ &$12$ & $500$\\
\hline
\hline
\end{tabular*}
\end{table} 

\subsection{CNN implementation}
\label{cnn:imp}

We use the same amount of instances used in Ref.~\cite{katzgraber:06}
with $100$ configurational overlaps at each temperature for each
instance.  Because the transition temperature with Gaussian disorder is
$T_c \approx 0.95$ \cite{marinari:98,katzgraber:05,katzgraber:06},
following Refs.~\cite{carrasquilla:17,carrasquilla:17b,tanaka:17} for
the training data, we label the convolutional overlaps with temperatures
above $0.95$ as ``1'' and those from temperatures below $0.95$ as ``0.''

The parameters for the architecture of the convolutional neural network
are listed in Tab.~\ref{tab:arch}. We inherit the structure with a
single layer from Ref.~\cite{tanaka:17}. All the parameters are
determined by extra validation sample sets, which are also generated
from Monte Carlo simulations.

\begin{table}[h]
\caption{CNN architecture, parameters, and hardware details. 
\label{tab:arch}
}
\begin{tabular*}{\columnwidth}{@{\extracolsep{\fill}}l l l  }
\hline
\hline
 Number of Layers & $1$ \\
\hline
 Channels in each layer& $5$\\
\hline
Filter size & $3\times3\times3$\\
\hline
stride & $2$\\
\hline
Activation function & ReLU\\
\hline
Optimizer & AdamOptimizer($10^{-4}$)\\
\hline
Batch size & $10^3$\\
\hline
Iteration & $10^4$\\
\hline
Software & TensorFlow (Python)\\
\hline 
Hardware & Lenovo x86 HPC cluster with a dual-GPU \\
         & NVIDIA Tesla K80 GPU and 128 GB RAM\\
\hline
\hline
\end{tabular*}
\end{table}	

Note that we use between $4000$ and $10000$ disorder instances for the
bimodal prediction data, which is approximately $1/3$ of the numerical
effort needed when estimating the phase transition directly via a
finite-scaling analysis of Monte Carlo data, as done for example in
Ref.~\cite{katzgraber:06}. As such, {\em pairing high-quality Monte
Carlo simulations with machine learning techniques can result in large
computational cost savings}.

\subsection{Data analysis}
\label{sec:analysis}
Because the configurational overlaps [Eq.~\eqref{overlaps}] include the
information about phases, we expect that different phases have different
overlap patterns similar to grid-like graphs. Therefore, in the region
of a specific phase, it is reasonable to expect that the classification
probability for the CNN to identify the phase correctly should be larger
than $50\%$. As such, it can be expected that when the classification
probability is $0.5$, the system is at the system-size-dependent critical
temperature. A thermodynamic estimate can then obtained via the
finite-size scaling method presented below.

\begin{figure}[h]
\includegraphics[width =\columnwidth]{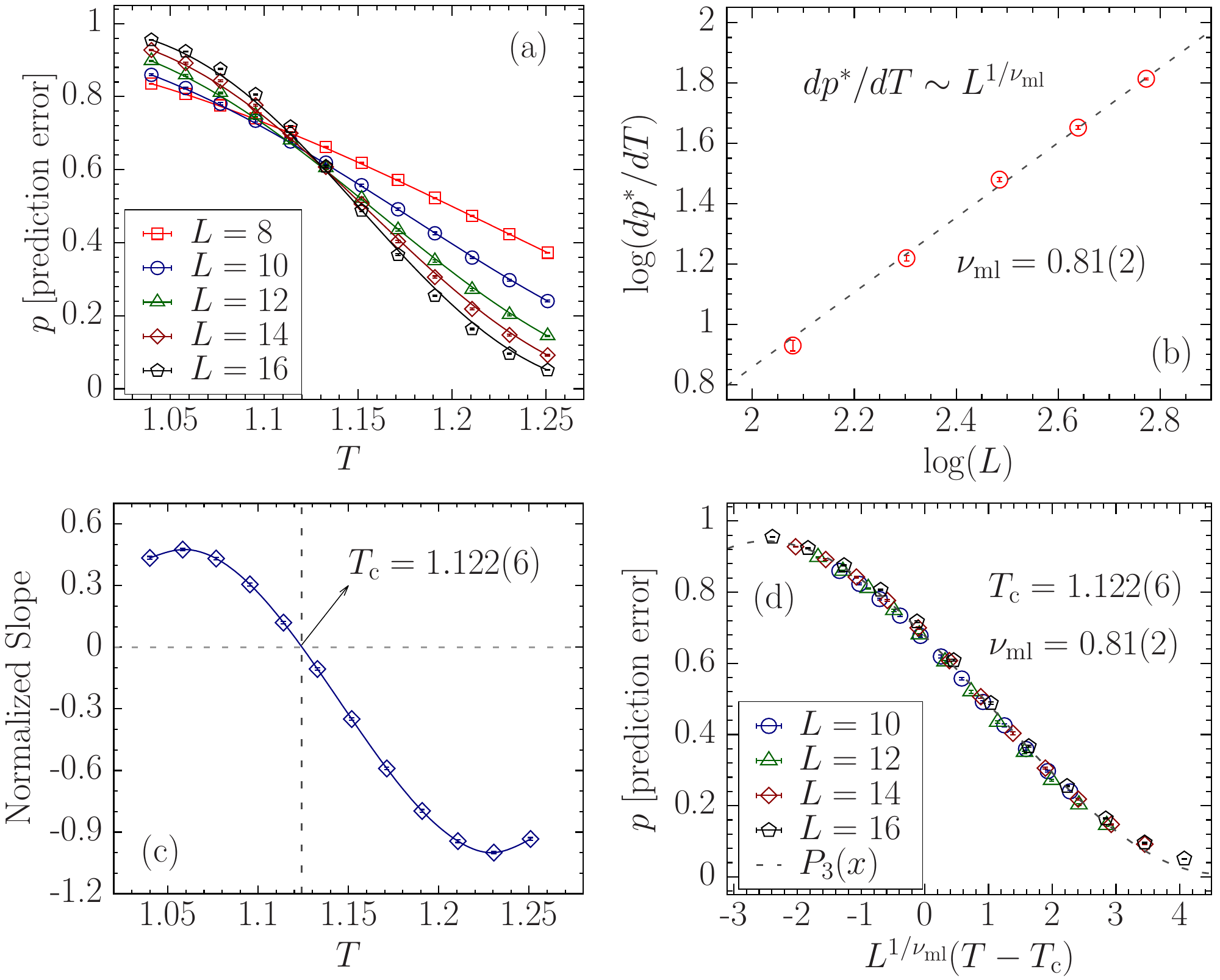}
\caption{
Classification probabilities for different linear system sizes $L$ as a
function of temperature $T$ for the prediction of the critical 
temperature of the bimodal Ising spin glass via a CNN trained with 
data from a Gaussian distribution.  
(a) Prediction probability for different system sizes $L$ near the phase
transition temperature. The different data sets cross at $T_{\rm c} \sim
1.122$. (b) Measurement of $\nu_{\rm ml}$ by performing a linear fit in
a double-logarithmic scale using the extremum points of the derivative
of the prediction error with respect to the temperature. (c) Estimate
of the critical temperature $T_{\rm c}$ using the coefficient of the
linear term in Eq.~\eqref{leading_order} (normalized to $1$) with
$L^{1/\nu_{\rm ml}}$ as the independent variable. The vertical dashed
line shows the temperature where the slope vanishes, which corresponds to
$T_{\rm c}$. (d) Finite-size scaling of the data using the
previously-estimated value of $\nu_{\rm  ml}$ and $T_{\rm c}$.  The data
collapse onto a universal curve indicating that the estimates are
accurate.}
\label{fig:good_data}
\end{figure}

Let us define the classification probability as a function of
temperature and system size: $p(T,L)$ which can be used as a
dimensionless quantity to describe the critical behavior. From the scaling
hypothesis, we expect $p(T,L)$ to have the following behavior in the
vicinity of the critical temperature $T_{\rm c}$:
\begin{equation}
\label{pt}
\langle p(T,L)\rangle = \tilde{F}\left[L^{1/\nu_{\rm ml}}\left(T-T_{\rm c}\right)\right],
\end{equation}
where the average is over disorder realizations. Note that the critical
exponent $\nu_{\rm ml}$ is different from the one calculated using
physical quantities. Due to the limited system sizes that we have
studied, finite-size scaling must be used to reliably calculate the
critical parameters at the thermodynamic limit. Assuming that we are
close enough to the critical temperature $T_{\rm c}$, the scaling
function $\tilde{F}$ in Eq.~\eqref{pt} can be expanded to a third-order
polynomial in $x=L^{1/\nu_{\rm ml}}\left(T-T_{\rm c}\right)$.
\begin{equation}
\label{leading_order}
\langle p(T,L)\rangle \sim p_0 + p_1x+p_2x^2+p_3x^3.
\end{equation}
First, we evaluate $\nu_{\rm ml}$ by noting that to the leading order in
$x$, the derivative of $\langle p(T,L)\rangle$ in
Eq.~\eqref{leading_order} with respect to temperature has the following
form:
\begin{align}
\label{nu-scaling}
\frac{d\langle p(T,L)\rangle}{dT} \sim L^{1/\nu_{\rm ml}}\left[p_1+2p_2L^{1/\nu_{\rm ml}}\left(T-T_{\rm c}\right)+\right.\nonumber\\
\left.3p_3L^{2/\nu_{\rm ml}}\left(T-T_{\rm c}\right)^2\right].
\end{align}

Therefore, the extremum point of $\frac{d\langle p(T,L)\rangle}{dT}$ scales as 
\begin{equation}
    \label{derivative-scaling}
    \frac{d\langle p(T,L)\rangle}{dT}|_{T=T^*}\sim L^{1/\nu_{\rm ml}}.
\end{equation}
A linear fit in a double-logarithmic scale then produces the value of
$\nu_{\rm ml}$ (slope of the straight line), which is subsequently used to
estimate $T_{\rm c}$. To do so, we turn back to Eq.~\eqref{leading_order}
where we realize that the coefficient of the linear term in $
L^{1/\nu_{\rm ml}}$ as the independent variable is proportional to
$(T-T_{\rm c})$ that changes sign at $T=T_{\rm c}$. Alternatively, we
can vary $T_{\rm c}$ until the data for all system sizes collapse onto a
common third-order polynomial curve. This is true because the scaling
function $\tilde{F}$ as a function of $L^{1/\nu_{\rm ml}}\left(T-T_{\rm
c}\right)$ is universal. The error bars can be computed using the
bootstrap method.

\begin{figure}[t!]
\includegraphics[width = 0.5\textwidth]{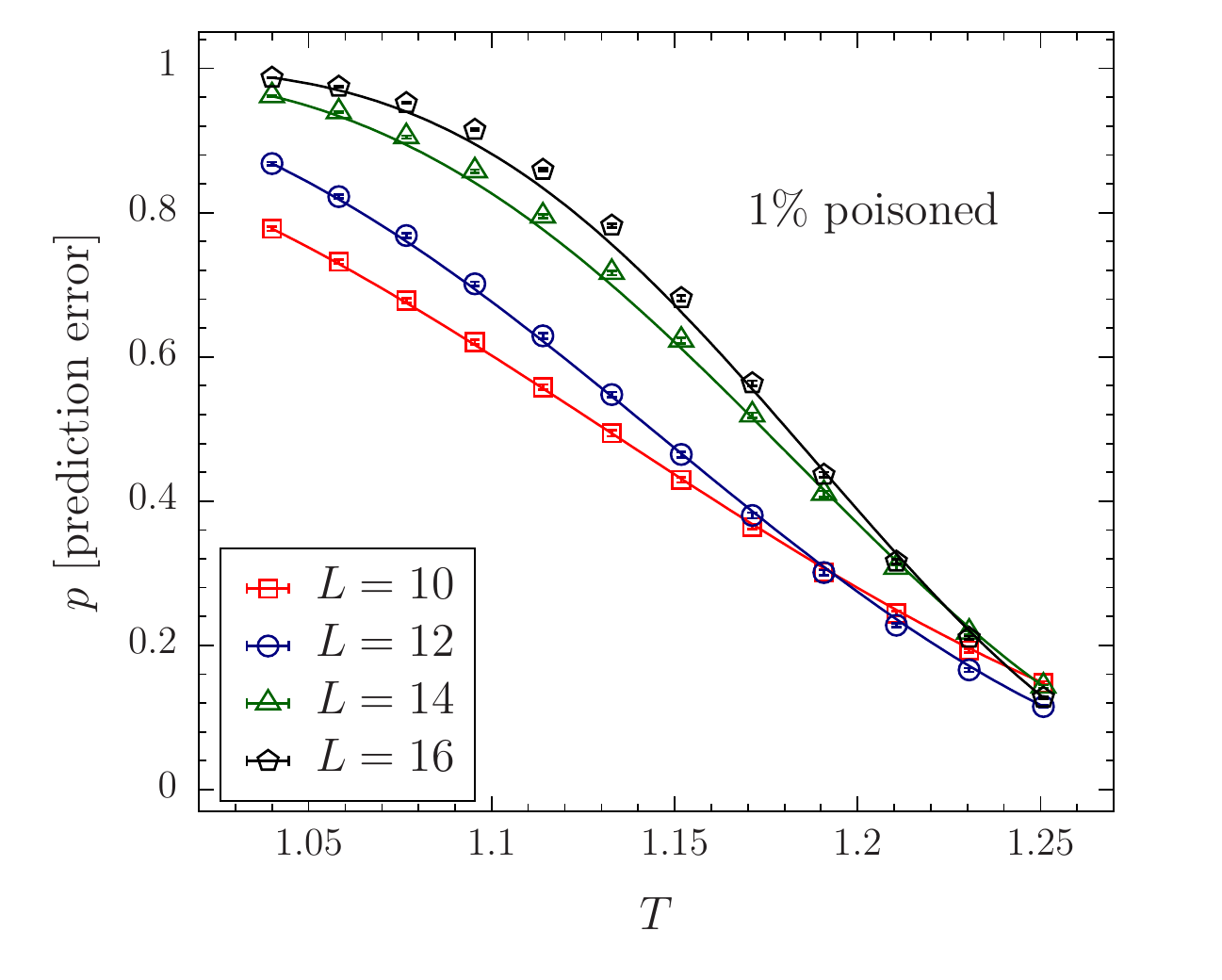}
\caption{
Classification probabilities for different system sizes $L$ for an Ising
spin glass with bimodal disorder. 1\% of the labels have been mixed on
average. There is no clear sign of the transition.  }
\label{fig:mixed_label}
\end{figure}

\section{Results using data without poisoning}

Figure \ref{fig:good_data} shows results from the CNN trained with
well-prepared (thermalized) data from a Gaussian distribution,
predicting the phase transition of data from a Bimodal disorder
distribution. Figure \ref{fig:good_data}(a) shows the prediction
probabilities for different linear system sizes $L$ as a function of
temperature $T$.  The curves cross the $p = 0.5$ line in the region of
the transition temperature for the bimodal Ising spin glass.  Figures
\ref{fig:good_data}(b) and \ref{fig:good_data}(c) show the estimates of
the exponent $\nu_{\rm ml}$ and the critical temperature $T_{\rm c}$,
respectively using the methods developed in Sec.~\ref{sec:analysis}.
The critical temperature $T_c = 1.122(6)$ is in good agreement with
previous estimates (see, for example, Ref.~\cite{katzgraber:06}).
Finally, in Fig.~\ref{fig:good_data}(d), the data points are plotted as
a function of the reduced variable $x=L^{1/\nu_{\rm ml}}\left(T-T_{\rm
c}\right)$ using the estimated values of the critical parameters. The
universality of the scaling curve underlines the accuracy of the
estimates.

\section{Results using poisoned training sets}

Although we have shown that the prediction from convolutional neural
network can be precise, we still need to test how poisoned data sets
impact the final prediction.  First, we randomly mix the classification
labels of the training sample with a probability of $1\%$, i.e., with a
training set of $100$ samples, this means only one mislabeled sample on
average. Then we train the network and use the same samples in the
prediction stage.  Compared to Fig.~\ref{fig:good_data},
Figure~\ref{fig:mixed_label} shows no clear sign of a phase transition.
This means that mislabeling a very small portion of the training data
can strongly affect the outcome. Given the hierarchical structure
of CNNs, errors can easily be amplified in propagation
\cite{Rumelhart:88,hecht-nielsen:92}, which is a possible explanation of
the observed behavior.

\begin{figure}[h]
\includegraphics[width =  0.46\textwidth]{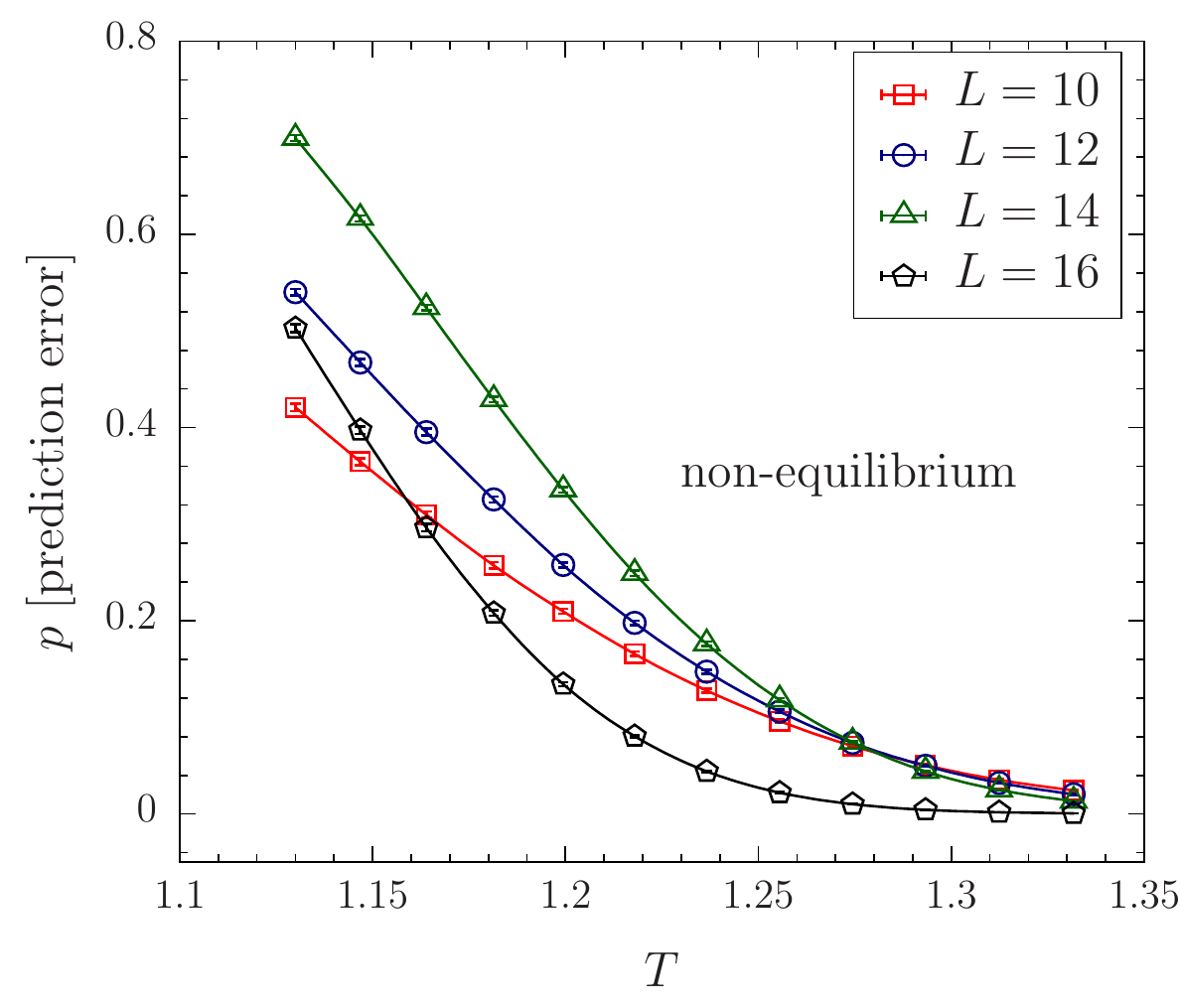}
\caption{
Classification probabilities for different system sizes $L$ for an Ising
spin glass with bimodal disorder. The Gaussian training data are not
thermalized.  There is no clear sign of a phase transition. }
\label{fig:non-equ}
\end{figure}

Finally, we test the effects of poorly prepared training data--in this
case, the training data are not properly thermalized. Figure
\ref{fig:non-equ} shows the prediction results using data with only
$50\%$ of the Monte Carlo sweeps needed for thermalization of the
Gaussian training samples. Although 50\% might seem extreme at first
sight, it is important to emphasize that thermalization times (as well
as time-to-solution) are typically distributed according to fat-tail
distributions \cite{steiger:15}. In general, users perform at least a
factor $2$ of additional thermalization to ensure most instances are
in thermal equilibrium. As in the case where the labels were mixed, a
transition cannot be clearly identified. This is strong indication that
the training data need to be carefully prepared.

We have also studied the effects of poorly-thermalized prediction data
paired with well-thermalized training data (not shown). In this case,
the impacts on the prediction probabilities are small but not
negligible.

\section{Discussion}

We have studied the effects of poisoned data sets when training CNNs to
detect phase transitions in physical systems. Our results show that good
training sets are a necessary requirement for good predictions. 
Small perturbations in the training set can lead to misleading results.

We do note, however, that we might not have selected the best parameters
for the CNN. Using cross-validation or bootstrapping might allow for a
better tuning of the parameters and thus improve the quality of the
predictions.  Furthermore, due to the large number of predictors,
overfitting is possible.  This, however, can be alleviated by the
introduction of penalty terms. Finally, the use of other activation
functions and optimizers can also impact the results.  This, together
with the sensitivity towards the quality of the training data that we find in
this work suggest that machine learning techniques should be used with
caution in physics applications. Garbage in, garbage out \ldots

\begin{acknowledgments}

We would like to thank Humberto Munoz Bauza and Wenlong Wang for
fruitful discussions.  This work is supported in part by the Office of
the Director of National Intelligence (ODNI), Intelligence Advanced
Research Projects Activity (IARPA), via MIT Lincoln Laboratory Air Force
Contract No.~FA8721-05-C-0002. The views and conclusions contained
herein are those of the authors and should not be interpreted as
necessarily representing the official policies or endorsements, either
expressed or implied, of ODNI, IARPA, or the U.S.~Government. The
U.S.~Government is authorized to reproduce and distribute reprints for
Governmental purpose notwithstanding any copyright annotation thereon.
We thank Texas A\&M University for access to their Terra cluster.

\end{acknowledgments}

\bibliography{refs}

\end{document}